# Design-Efficiency in Security

Ender Yüksel
Hanne Riis Nielson
Flemming Nielson

Technical University of Denmark

# Table of Contents

# List of Tables

# List of Figures

## Foreword

In this document, we present our applied results on balancing security and performance using a running example, which is based on sensor networks. These results are forming a basis for a new approach to balance security and performance, and therefore provide *design-efficiency* of key updates.

We employ probabilistic model checking approach and present our modelling and analysis study using PRISM model checker.

# 1. Running Example: Hotel Security Management

In this section, we describe the running example that we will use for showing how the design-efficiency approach works. This example is taken from [11YNN+].

## 1.1. Scenario

In this scenario, we focus on a commercial building automation application, specifically *hotel security management*. We consider a system where we use wireless sensors embedded in door locking cards which allow remote cancellation of cards, remote report of door lock status and door ajar alarms, etc.

The technical details of such a system in this example includes a maximum number of 50 devices in the network, with the aim of keeping the network at its maximum size as much as possible. Besides, there exists certain goals such that:
- stolen or broken cards will be replaced in two days on average,
- each device is expected to have non-stop operation for a year on average,
- each device sending one message a day on average, and
- the probability of a key compromise by either leaving devices or sent messages is one in ten thousand.

We use six different key update methods that were defined in [11YNN+]:
- Leave-based Key Update (LB)
- Join-based Key Update (JB)
- Join-Leave-based Key Update (JLB)
- Time-based key update (TB).
- Message-based Key Update (MB)
- Hybrid Key Update excluding MB (Hy/MB)

We will only use the abbreviations of these strategies, written in parenthesis above, in the rest of this report.

There are two criteria that need to be optimized. We can define these in terms of requirements such that:

**R1:** *The key compromise probability should be less than a specific percentage in the long run*.

**R2:** *Maximum allowed number of key updates is should be less than a specific number per year*.

In the following sections, we will detail the two requirements above.

## 1.2. Quantitative Security Analyses

Previous quantitative security analyses described in [11YNN+] were defining the details of the security analyses using probabilistic models and paving the way for the design-efficiency curves that we will use in this report. Below in Figure 1, we present sample graphical results of [11YNN+] in order to give a clue on the types of data that we can make use of. An automated tool that assists the decision of key update strategy using the type of graphics (and surely the data) in Figure 1 is available on [12YNNa].

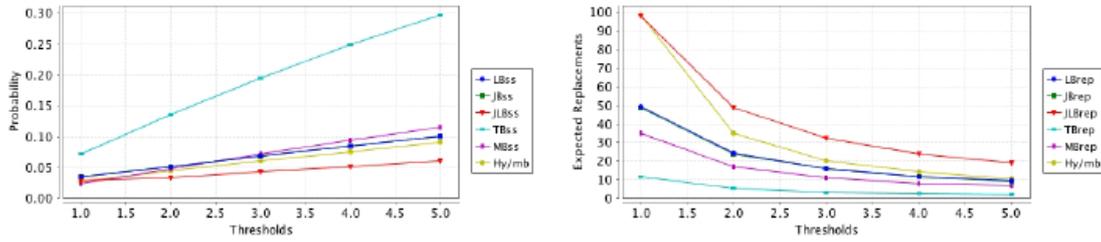

Figure 1: Previous Quantitative Results. Left: Key compromise probability in the long run, Right: Number of key updates in one year of time.

## 2. Design-Efficiency Approach

In this section, we describe our method for generating design-efficiency curves on our running example.

From [12YNN], we already know that the risk of key compromise tends to stabilise by time, and this stabilisation period depends on the chosen key update strategies, the chosen parameter value (i.e. threshold), and the network dynamics. We also know that, until stabilisation, there could be fluctuations in the risk, such that the maximum risk is often reached within this period. Therefore, we subdivide our two criteria – Risk and Cost – depending on the stabilisation.

The formal models and input values of the analysis are available in the Appendix.

### 2.1. Risk of Key Compromise

**Key compromise in the long run.** We start by computing the steady-state probabilities for a (moderate) set of thresholds. The aim of this step is to find the set of thresholds for each key update method that satisfies R1. In the last column of Table 1, we have presented numerical results of probabilistic model checking, which has the meaning: *the risk of the key being compromised in the long run*.

**Key compromise at a specific time.** We continue with computing the transient probabilities for the same set of thresholds. The aim of this step is to see if the peak points of risk exceed the requirement R1 significantly or not. In Table 1, we have presented the results for our running example, which has the meaning: *the maximum risk of the key being compromised*. In the cases where the maximum and the average risk are the same in the table, the difference (or deviation from average) is relatively small.

A mathematical way of computing the maximum risk of key compromise is described in [12YNNN] for LB key updates, and a tool implemented in MATLAB that automates this procedure is provided. In this running example, we have found the maximum risk manually from the probabilistic model checking results of PRISM [06HKM+].

Table 1: Risk of Key Compromise

|       | Threshold | RISK (max) | RISK (average) |
|-------|-----------|------------|----------------|
| **LB** | 1 | 0.035 | 0.035 |
|       | 2 | 0.052 | 0.052 |
|       | 3 | 0.069 | 0.069 |
|       | 4 | 0.085 | 0.085 |
|       | 5 | 0.101 | 0.101 |
| **JB** | 1 | 0.035 | 0.035 |
|       | 2 | 0.052 | 0.052 |
|       | 3 | 0.069 | 0.069 |
|       | 4 | 0.087 | 0.085 |
|       | 5 | 0.104 | 0.101 |
| **JLB** | 1 | 0.029 | 0.029 |
|       | 2 | 0.034 | 0.034 |
|       | 3 | 0.044 | 0.044 |
|       | 4 | 0.052 | 0.052 |
|       | 5 | 0.062 | 0.061 |
| **TB** | 1 | 0.074 | 0.072 |
|       | 2 | 0.139 | 0.137 |
|       | 3 | 0.259 | 0.196 |
|       | 4 | 0.36  | 0.249 |
|       | 5 | 0.443 | 0.298 |
| **MB** | 1 (500)  | 0.029 | 0.025 |
|       | 2 (1000) | 0.064 | 0.048 |
|       | 3 (1500) | 0.098 | 0.072 |
|       | 4 (2000) | 0.139 | 0.094 |
|       | 5 (2500) | 0.18  | 0.115 |
| **Hy/MB** | 1 | 0.027 | 0.027 |
|       | 2 | 0.044 | 0.044 |
|       | 3 | 0.062 | 0.060 |
|       | 4 | 0.081 | 0.076 |
|       | 5 | 0.099 | 0.092 |

## 2.2. Cost of Key Updates

**Number of key updates before stabilisation:** In this criterion, we compute the expected number of key updates *before* the risk of key compromise gets stabilised. Obviously, we need to be able to figure out the time point that the risk stabilises. Then, we normalize the costs to get the monthly numbers, such that we can see the expected number of key updates per month no matter how long or short the stabilisation period is.

As in the case of maximum risk above, a mathematical way of computing the stabilisation point is described in [12YNNN] for LB key updates, and a tool was implemented in MATLAB that automates this procedure. In this running example, we have found the stabilisation points manually.

The results for our running example are presented in Table 2, in the same fashion with the results on the risk of key compromise.

**Number of key updates after stabilisation:** In this criterion, we compute the expected number of key updates *after* the risk of key compromise gets stabilised. We do this similar to the process we use for pre-stabilisation results. However, we need to define an observation period, and we defined this period as 12 months. In practice, we compute the expected number of key updates until the month $S$+12, where $S$ is the month of stabilisation. Then, we subtract the expected number of key updates in the stabilisation period, which we know from the previous step. In the end, we provide monthly average results by simply dividing to number of observation months, that is 12.

The results for our running example are presented in the last column of Table 2, in the same fashion with the results on the risk of key compromise.

Table 2: Cost of Key Updates

|     | Threshold | COST per month (before stabilisation) | COST per month (after stabilisation) |
| --- | --- | --- | --- |
| LB  | 1 | 4.089 | 4.088 |
|     | 2 | 1.919 | 2.044 |
|     | 3 | 1.196 | 1.363 |
|     | 4 | 0.835 | 1.022 |
|     | 5 | 0.618 | 0.817 |
| JB  | 1 | 3.817 | 4.088 |
|     | 2 | 1.658 | 2.044 |
|     | 3 | 0.938 | 1.363 |
|     | 4 | 0.801 | 1.022 |
|     | 5 | 0.591 | 0.817 |
| JLB | 1 | 8.041 | 8.175 |
|     | 2 | 3.847 | 4.088 |
|     | 3 | 2.301 | 2.725 |
|     | 4 | 1.859 | 2.044 |
|     | 5 | 1.408 | 1.635 |
| TB  | 1 | 0.755 | 1.000 |

| | 2 | 0.438 | 0.500 |
|---|---|---|---|
| | 3 | 0.327 | 0.333 |
| | 4 | 0.246 | 0.250 |
| | 5 | 0.196 | 0.200 |
| **MB** | 1 (500) | 2.960 | 2.985 |
| | 2 (1000) | 1.482 | 1.495 |
| | 3 (1500) | 0.931 | 0.993 |
| | 4 (2000) | 0.742 | 0.749 |
| | 5 (2500) | 0.592 | 0.591 |
| **Hy/MB** | 1 | 7.920 | 8.232 |
| | 2 | 2.562 | 2.959 |
| | 3 | 1.515 | 1.736 |
| | 4 | 1.066 | 1.221 |
| | 5 | 0.782 | 0.940 |

**A note on the stabilisation:** Even though all key update strategies used in this report are expected to be stabilised in terms of risk of key compromise, for some of the cases this period is fairly long. Therefore, we have assumed the stabilisation point of such cases to be 10 years. Precise data is given in Table 3, such that the month where the deviation of the risk drops below 0.001 is assumed to be the month of stabilisation.

**Table 3: Stabilisation Months**

| Strategy | Threshold | Stabilisation (month) | Strategy | Threshold | Stabilisation (month) |
|---|---|---|---|---|---|
| **LB** | 1 | 1 | **TB** | 1 | 2 |
| | 2 | 2 | | 2 | 8 |
| | 3 | 2 | | 3 | 77 |
| | 4 | 2 | | 4 | 120 |
| | 5 | 2 | | 5 | 120 |
| **JB** | 1 | 1 | **MB** | 1 | 22 |
| | 2 | 1 | | 2 | 54 |
| | 3 | 1 | | 3 | 11 |
| | 4 | 2 | | 4 | 120 |
| | 5 | 2 | | 5 | 120 |
| **JLB** | 1 | 2 | **Hy/MB** | 1 | 1 |
| | 2 | 1 | | 2 | 1 |
| | 3 | 1 | | 3 | 2 |
| | 4 | 2 | | 4 | 3 |
| | 5 | 2 | | 5 | 3 |

## 2.3. Design-Efficiency Curves

In this section, we introduce the *design-efficiency curves* on the running example. We present the same style of graphics, i.e. **Risk** of key compromise on the vertical axis, and **Cost** of key update on the horizontal axis. The risk is given as a percentage, whereas the cost is given as monthly number of updates. In the first two subsections, the emphasis is on the stabilisation of the network. In the last two subsections, the emphasis is on the combination of curves.

### 2.3.1. Until Stabilisation

In the period until the stabilisation of the network - in fact the risk of key compromise in the network - fluctuations are expected in the risk and therefore there will be points where the risk is maximum, and points where the risk is significantly deviating from the average. Therefore this period is important for the designers, who cannot tolerate temporary peaks that exceed a certain level of risk, or significantly deviate from the risk in the long run. Similarly, this period is important for the designers who cannot tolerate excessive key updates that could take place in this period.

Below we present the design-efficiency curves that we produced for the running example for the period before stabilisation. The risk value in the vertical axis should be interpreted as the (percentage of) maximum risk of key compromise during the period until stabilisation.

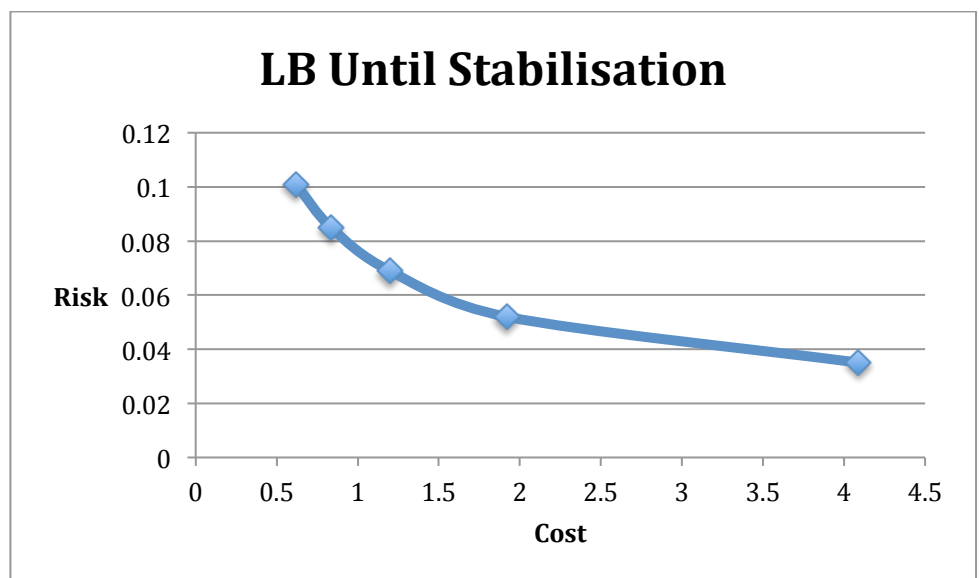

Figure 2: Design-Efficiency Curve of LB Key Update until stabilisation

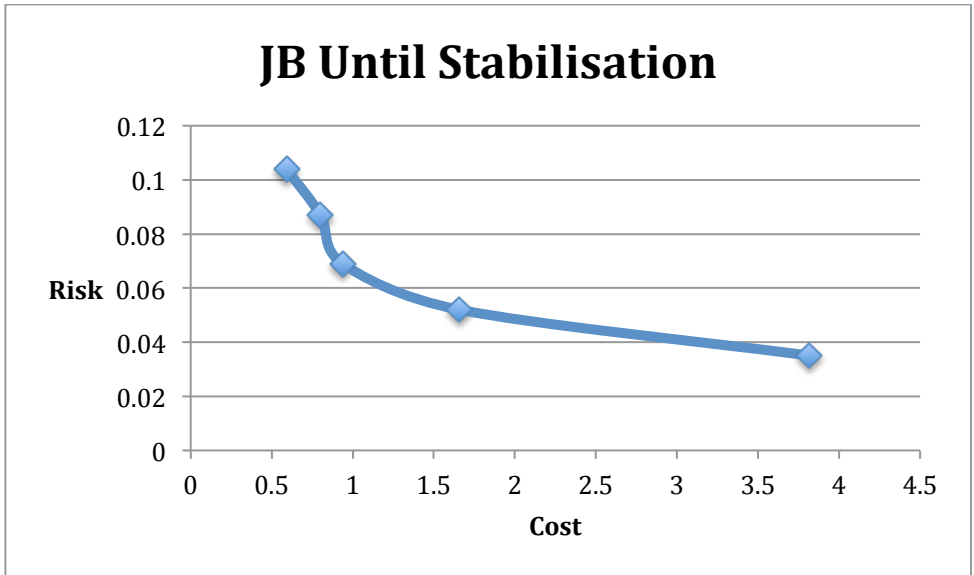

**Figure 3: Design-Efficiency Curve of JB Key Update until stabilisation**

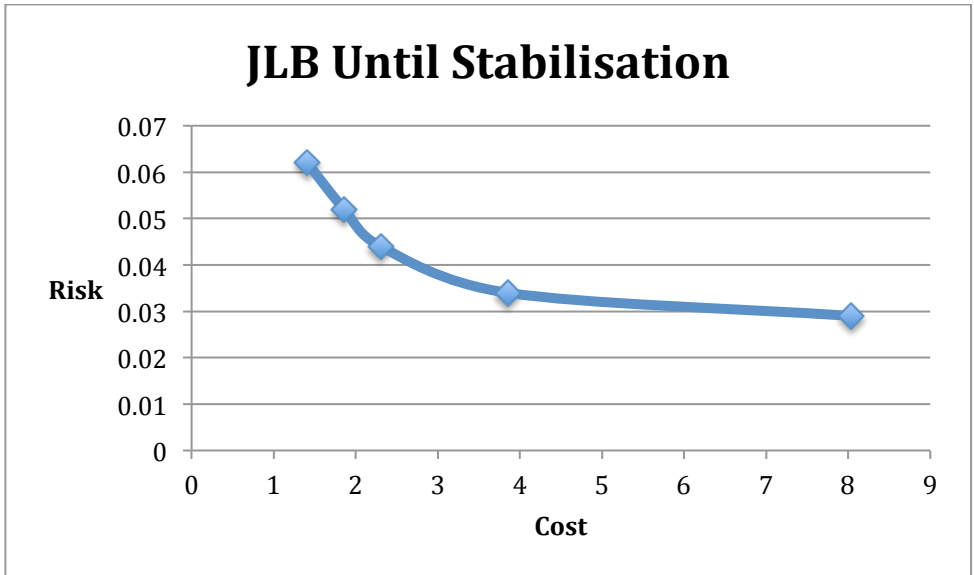

**Figure 4: Design-Efficiency Curve of JLB Key Update until stabilisation**

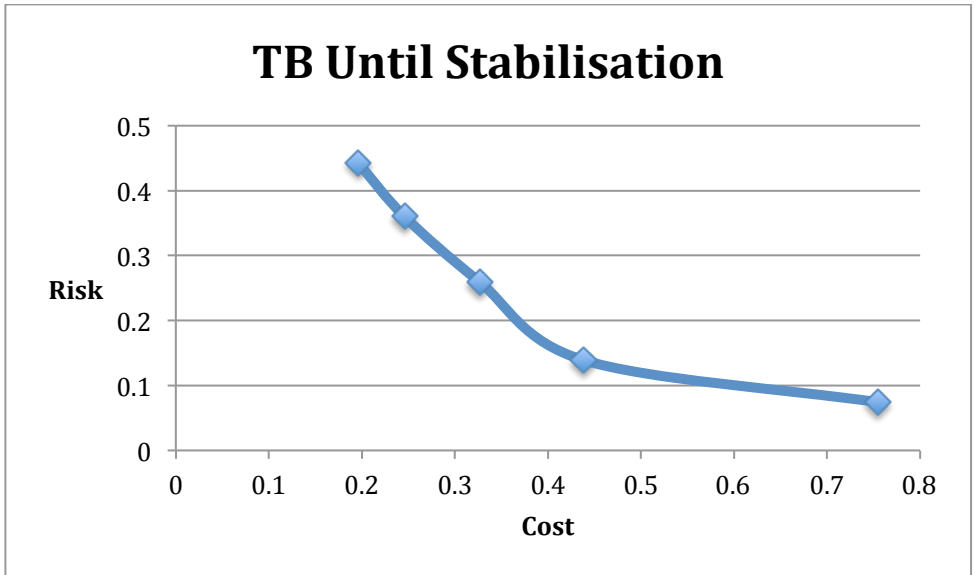

**Figure 5: Design-Efficiency Curve of TB Key Update until stabilisation**

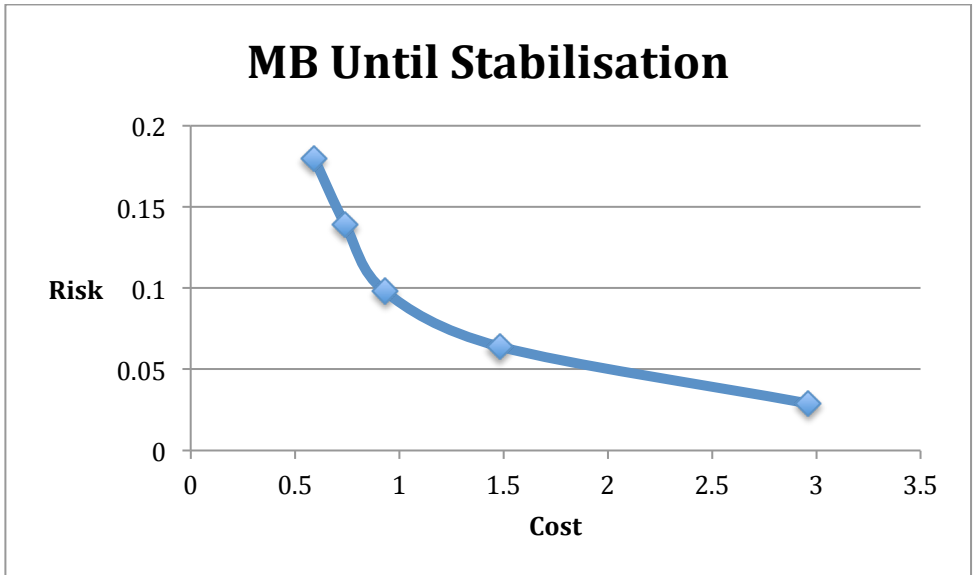

**Figure 6: Design-Efficiency Curve of MB Key Update until stabilisation**

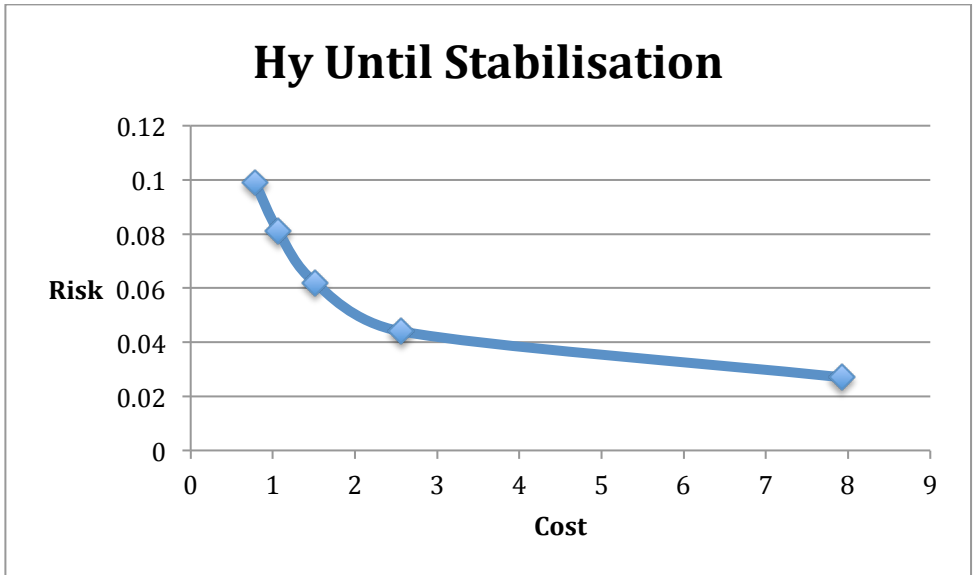

Figure 7: Design-Efficiency Curve of Hy Key Update until stabilisation

### 2.3.2 After Stabilisation

In the period after the stabilisation of the network, fluctuations are not expected in the risk and therefore the notions of maximum risk and minimum risk are not relevant anymore.

Below we present the design-efficiency curves that we produced for the running example for the period after stabilisation.

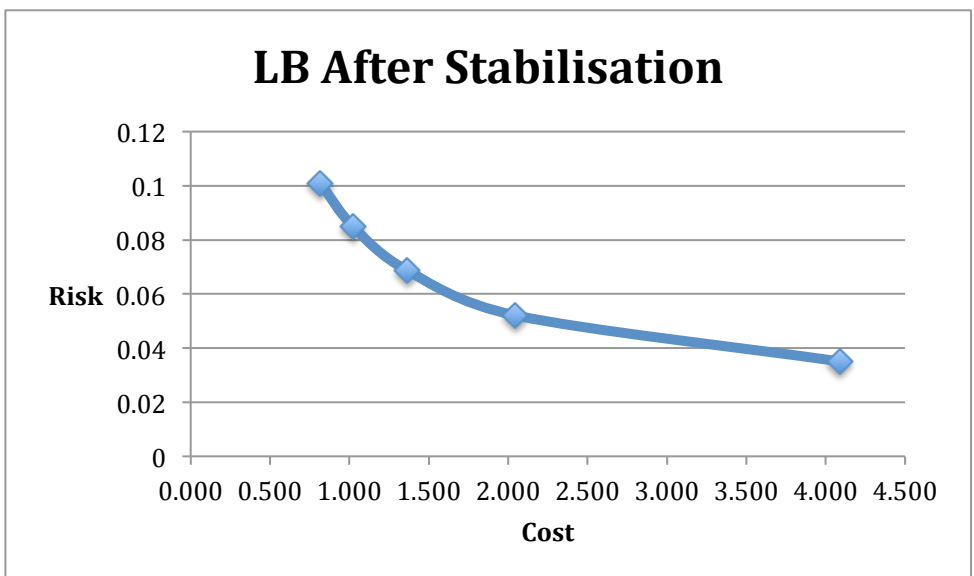

Figure 8: Design-Efficiency Curve of LB Key Update after stabilisation

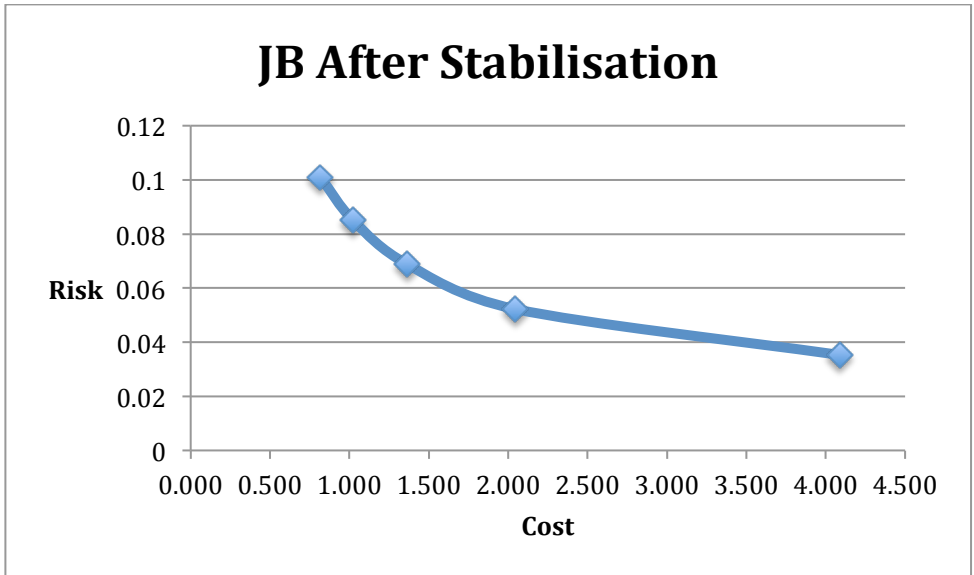

**Figure 9:** Design-Efficiency Curve of JB Key Update after stabilisation

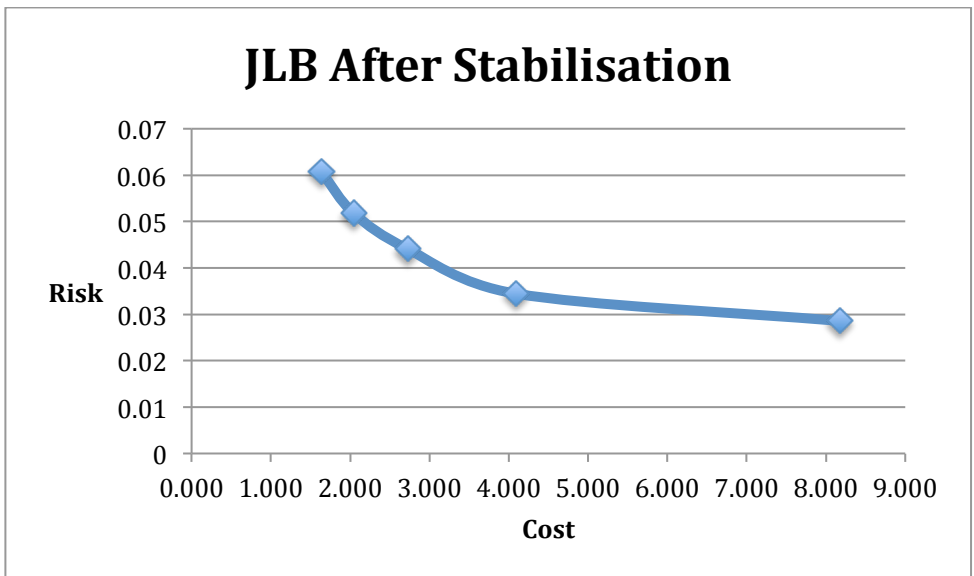

**Figure 10:** Design-Efficiency Curve of JLB Key Update after stabilisation

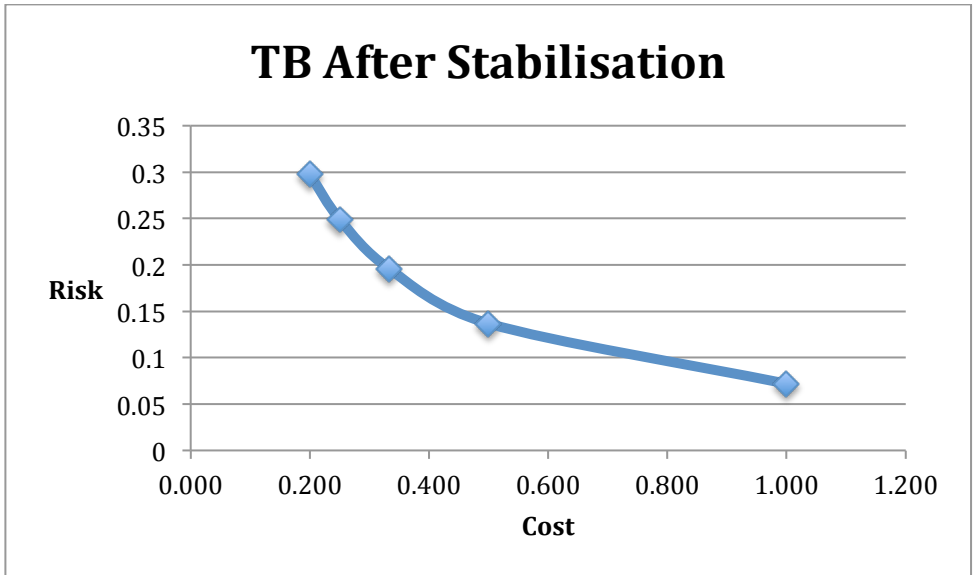

**Figure 11: Design-Efficiency Curve of TB Key Update after stabilisation**

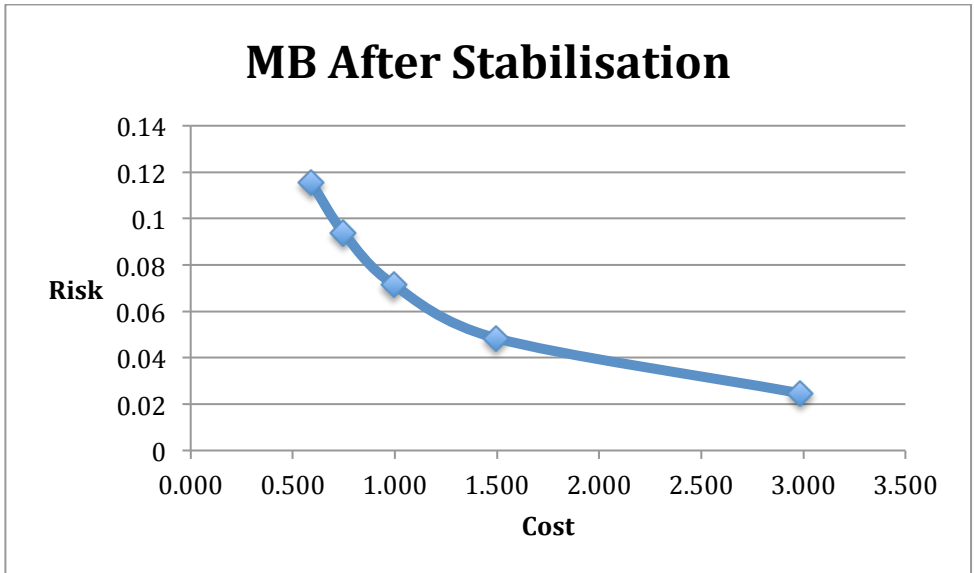

**Figure 12: Design-Efficiency Curve of MB Key Update after stabilisation**

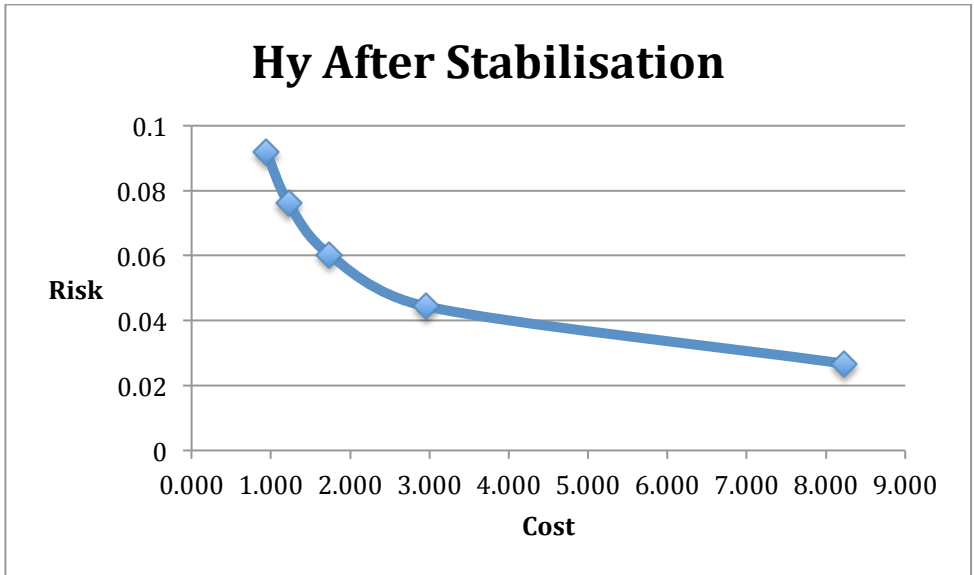
**Figure 13: Design-Efficiency Curve of Hy Key Update after stabilisation**

### 2.3.3. Combined Curves: Before and After Stabilisation

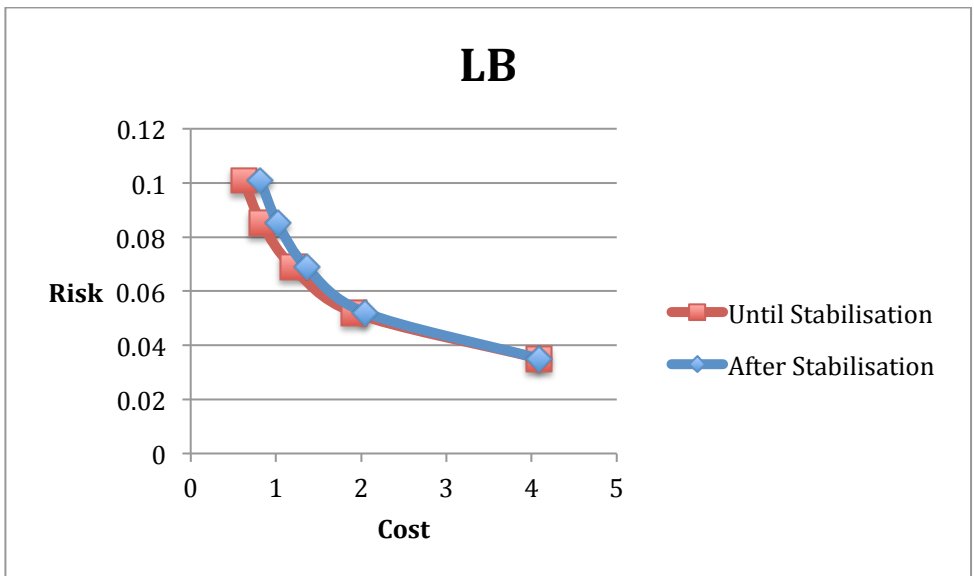
**Figure 14: Combined Design-Efficiency Curve for LB**

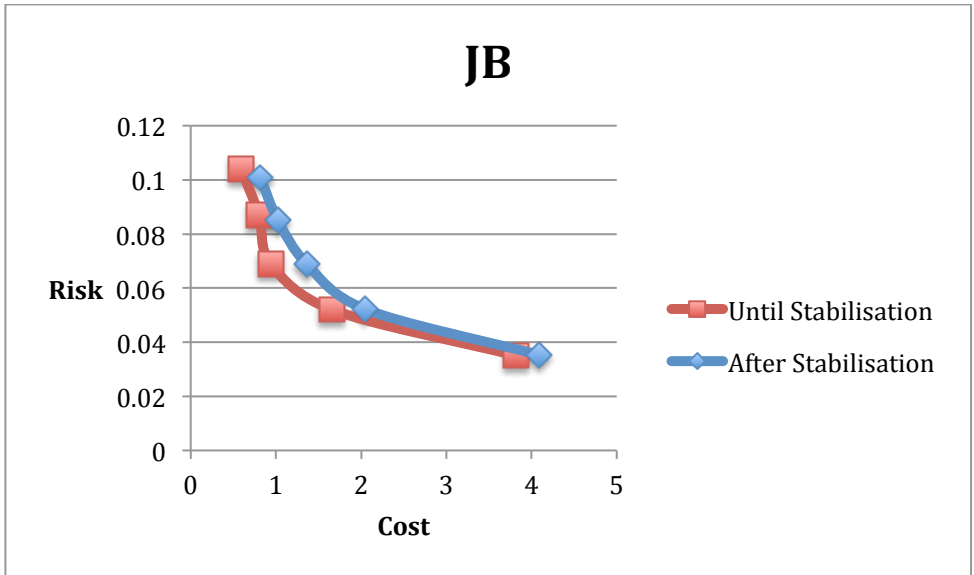

**Figure 15: Combined Design-Efficiency Curve for JB**

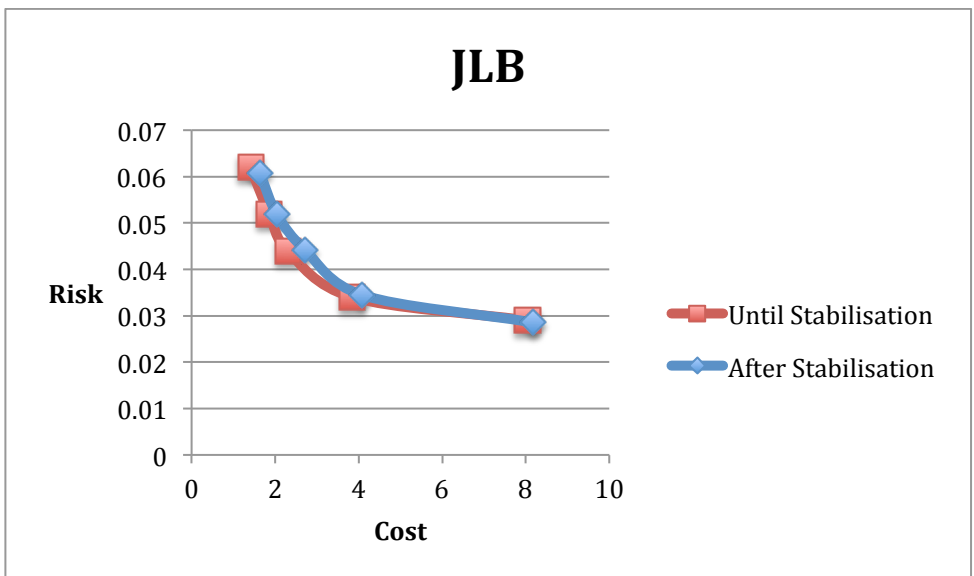

**Figure 16: Combined Design-Efficiency Curve for JLB**

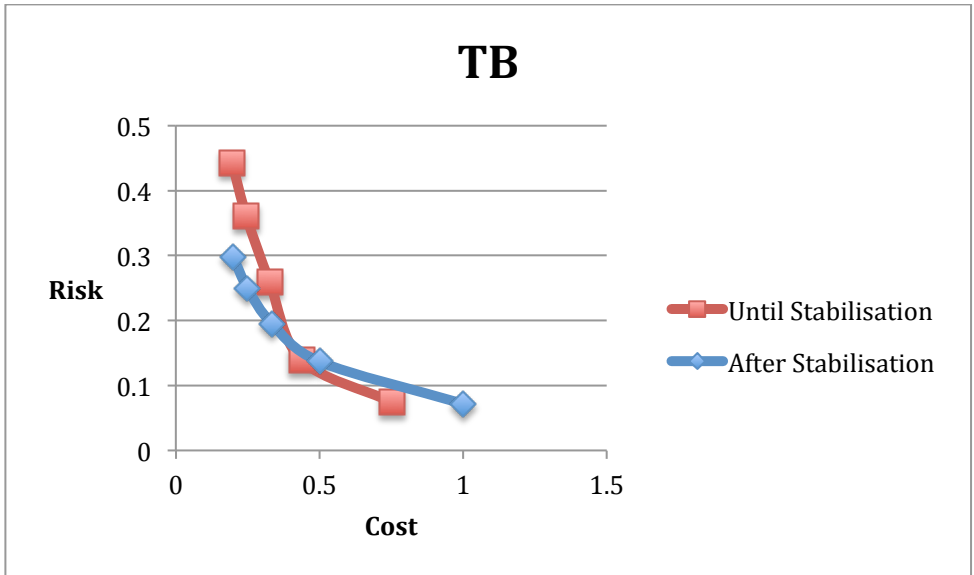

**Figure 17: Combined Design-Efficiency Curve for TB**

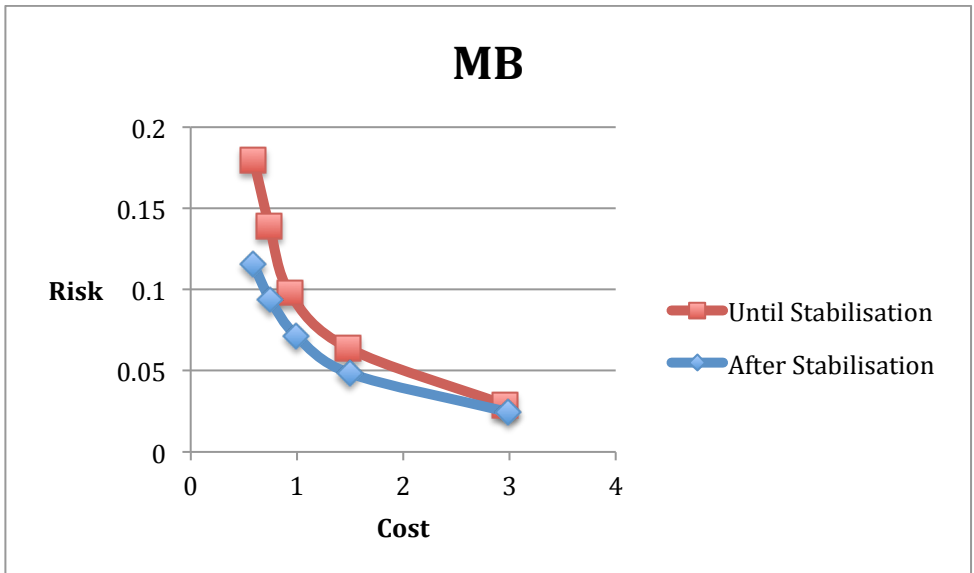

**Figure 18: Combined Design-Efficiency Curve for MB**

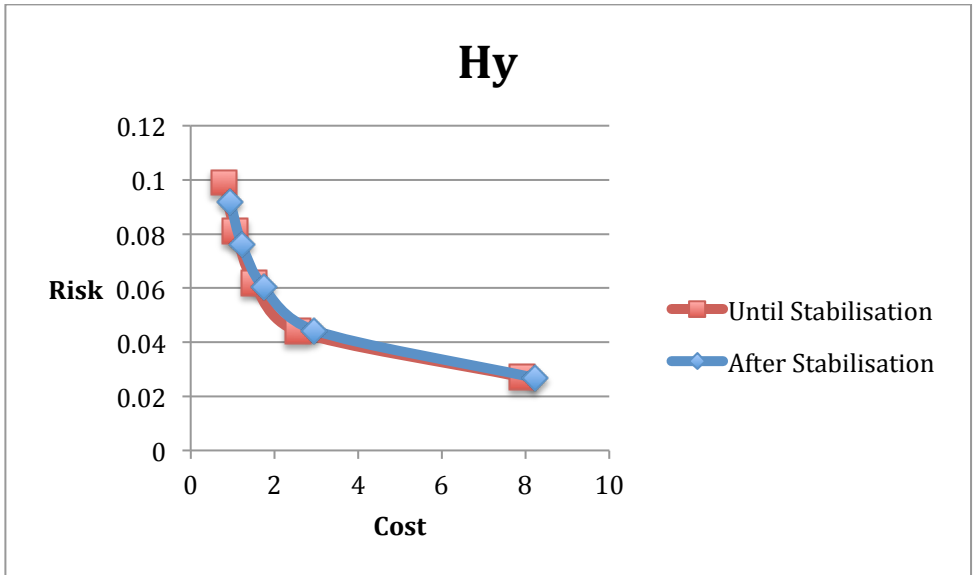

Figure 19: Combined Design-Efficiency Curve for Hy

### 2.3.4. Combined Curves: All Strategies

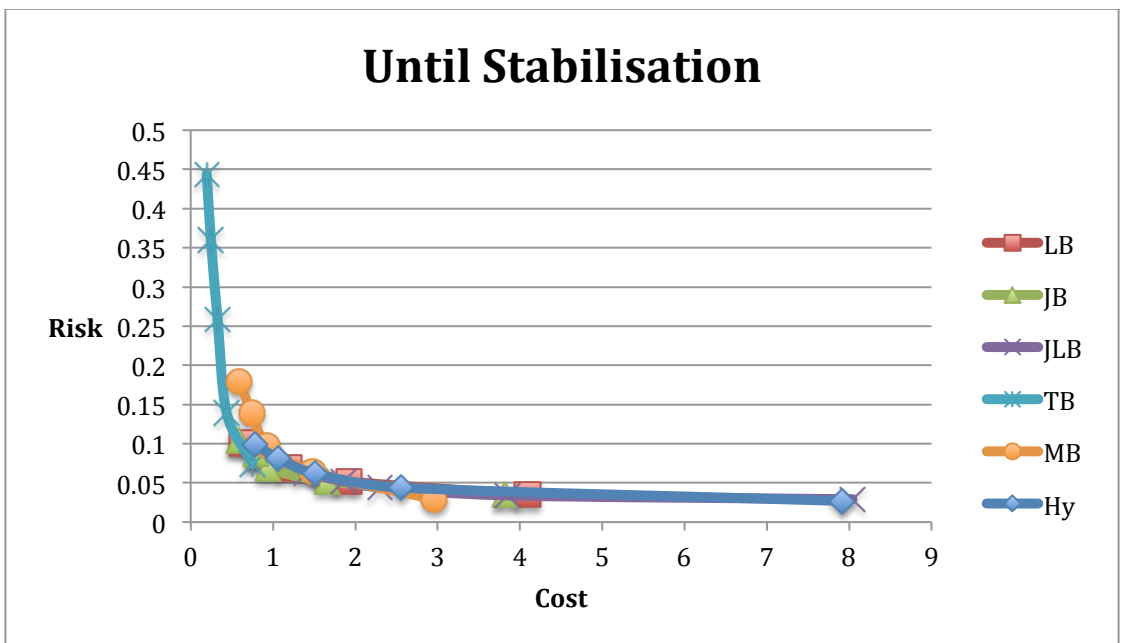

Figure 20: Combined Design-Efficiency Curves until stabilisation

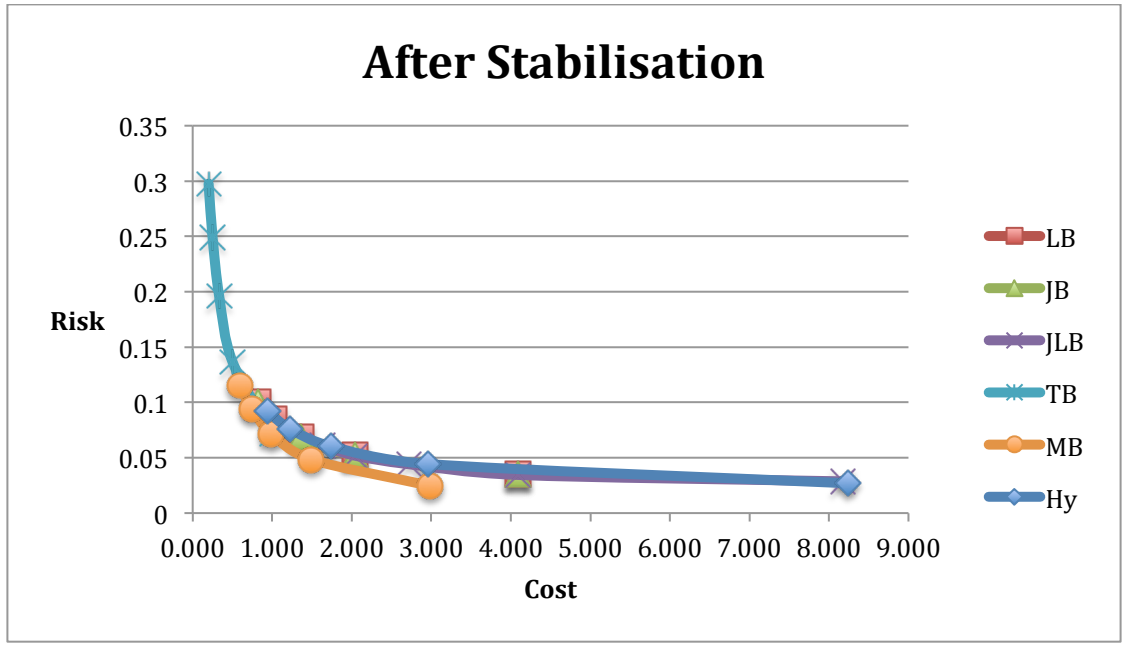

Figure 21: Combined Design-Efficiency Curves after stabilisation

## 3. Conclusion

In this technical report, we have introduced the design-efficiency curves approach by providing graphical results covering several key update strategies in a running example. We have provided all the necessary information for replicating the experiments including models, property specifications, and input parameters.

# APPENDIX

We present the PRISM models, property specification, and the input parameters that we used in producing the results in this report. Therefore, all the experiments can be easily replicated using the PRISM model checker.

## A. PRISM Model for LB Key Update

```
ctmc
// time unit: 1 day

const int N; // threshold for number of leaves

const int    Max;       // Maximum number of devices
const double R_join;    // Rate of join per device
const double R_leave;   // Rate of leave per device
const double R_message; // Rate of message per device
const double P_comp;    // Risk of key leakage per device

module DEVICES
 Size: [0..Max] init Max;

 [join]    Size<Max -> R_join*(Max-Size):      (Size'=Size+1);
 [leave]   Size>0   -> R_leave*(1-P_comp)*Size: (Size'=Size-1);
 [leaveC]  Size>0   -> R_leave*P_comp*Size:     (Size'=Size-1);
 [leaveR]  Size>0   -> R_leave*Size:            (Size'=Size-1);
 [message] Size>0   -> R_message*(1-P_comp)*Size:           true;
 [messageC] Size>0  -> R_message*P_comp*Size:       true;
endmodule

module COORDINATOR
 Comp: bool init false;
 C_leave: [0..N] init 0;

 [join]    true -> true;
 [leave]   C_leave<N-1 -> (C_leave'=C_leave+1);
 [leaveC]  C_leave<N-1 -> (C_leave'=C_leave+1) & (Comp'=true);
 [leaveR]  C_leave=N-1-> (C_leave'=0) & (Comp'=false);
 [message] true  ->         true;
 [messageC] true ->        (Comp'=true);
endmodule

rewards "Replacements"
  [leaveR] true: 1;
endrewards
```

## B. PRISM Model for JB Key Update

```
ctmc
// time unit: 1 day

const int J; // threshold for number of joins

const int    Max;       // Maximum number of devices
const double R_join;    // Rate of join per device
const double R_leave;   // Rate of leave per device
const double R_message; // Rate of message per device
const double P_comp;    // Risk of key leakage per device

module DEVICES
```

```
 Size: [0..Max] init Max;

 [join]    Size<Max  -> R_join*(Max-Size):     (Size'=Size+1);
 [joinR]   Size<Max  -> R_join*(Max-Size):     (Size'=Size+1);
 [leave]   Size>0    -> R_leave*(1-P_comp)*Size: (Size'=Size-1);
 [leaveC]  Size>0    -> R_leave*P_comp*Size:    (Size'=Size-1);
 [message] Size>0    -> R_message*(1-P_comp)*Size:          true;
 [messageC] Size>0   -> R_message*P_comp*Size:   true;
endmodule

module COORDINATOR
 Comp: bool init false;
 C_join: [0..J] init 0;

 [join]    C_join<J-1 -> (C_join'=C_join+1);
 [joinR]   C_join=J-1 -> (C_join'=0) & (Comp'=false);
 [leave]   true -> true;
 [leaveC]  true -> (Comp'=true);
 [message] true  ->           true;
 [messageC] true ->           (Comp'=true);
endmodule

rewards "Replacements"
  [joinR] true:  1;
endrewards
```

## C. PRISM Model for JLB Key Update

```
ctmc
// time unit: 1 day

const int JL; // threshold for number of leave and joins

const int    Max;      // Maximum number of devices
const double R_join;    // Rate of join per device
const double R_leave;   // Rate of leave per device
const double R_message; // Rate of message per device
const double P_comp;    // Risk of key leakage per device

module DEVICES
 Size: [0..Max] init Max;

 [join]    Size<Max  -> R_join*(Max-Size):     (Size'=Size+1);
 [joinR]   Size<Max  -> R_join*(Max-Size):     (Size'=Size+1);
 [leave]   Size>0    -> R_leave*(1-P_comp)*Size: (Size'=Size-1);
 [leaveC]  Size>0    -> R_leave*P_comp*Size:    (Size'=Size-1);
 [leaveR]  Size>0    -> R_leave*Size:           (Size'=Size-1);
 [message] Size>0    -> R_message*(1-P_comp)*Size:          true;
 [messageC] Size>0   -> R_message*P_comp*Size:    true;
endmodule

module COORDINATOR
 Comp: bool init false;
 C_joinleave: [0..JL] init 0;

 [join]    C_joinleave<JL-1 -> (C_joinleave'=C_joinleave+1);
 [joinR]   C_joinleave=JL-1 -> (C_joinleave'=0) & (Comp'=false);
 [leave]   C_joinleave<JL-1 -> (C_joinleave'=C_joinleave+1);
 [leaveC]  C_joinleave<JL-1 -> (C_joinleave'=C_joinleave+1) & (Comp'=true);
 [leaveR]  C_joinleave=JL-1 -> (C_joinleave'=0) & (Comp'=false);
 [message] true  ->           true;
 [messageC] true ->           (Comp'=true);
endmodule

rewards "Replacements"
```

```
  [leaveR] true:  1;
  [joinR] true:  1;
endrewards
```

## D. PRISM Model for TB Key Update

```
ctmc
// time unit: 1 day

const int M; // number of months between resets

const int k;
const double mean = 30*M;

const int    Max;       // Maximum number of devices
const double R_join;    // Rate of join per device
const double R_leave;   // Rate of leave per device
const double R_message; // Rate of message per device
const double P_comp;    // Risk of key leakage per device

module DEVICES
 Size: [0..Max] init Max;

 [leave]  Size>0   -> R_leave*(1-P_comp)*Size: (Size'=Size-1);
 [leaveC] Size>0   -> R_leave*P_comp*Size:     (Size'=Size-1);
 [join]   Size<Max -> R_join*(Max-Size):       (Size'=Size+1);
 [message] Size>0  -> R_message*(1-P_comp)*Size:             true;
 [messageC] Size>0 -> R_message*P_comp*Size:    true;
 [reset]  true     -> 1:                        true;
endmodule

module COORDINATOR
 Comp: bool init false;
 i : [1..k+1];

 [join]   true ->           true;
 [leave]  true ->           true;
 [leaveC] true ->           (Comp'=true);
 []        i < k -> k/mean :       (i'=i+1);
 [message] true ->          true;
 [messageC] true ->         (Comp'=true);
 [reset] i = k -> k/mean : (i'=1) & (Comp'=false);
endmodule

rewards "Replacements"
  [reset] true:  1;
endrewards
```

## E. PRISM Model for MB Key Update

```
ctmc
// time unit: 1 day

const int MSG; // threshold for number of messages

const int    Max;       // Maximum number of devices
const double R_join;    // Rate of join per device
const double R_leave;   // Rate of leave per device
const double R_message; // Rate of message per device
const double P_comp;    // Risk of key leakage per device

module DEVICES
```

```
 Size: [0..Max] init Max;

 [join]     Size<Max  -> R_join*(Max-Size):     (Size'=Size+1);
 [leave]    Size>0    -> R_leave*(1-P_comp)*Size: (Size'=Size-1);
 [leaveC]   Size>0    -> R_leave*P_comp*Size:    (Size'=Size-1);
 [message]  Size>0    -> R_message*(1-P_comp)*Size:          true;
 [messageC] Size>0    -> R_message*P_comp*Size:          true;
 [messageR] Size>0    -> R_message*Size:          true;
endmodule

module COORDINATOR
 Comp: bool init false;
 C_msg: [0..MSG] init 0;

 [join]    true -> true;
 [leave]   true -> true;
 [leaveC]  true -> (Comp'=true);
 [message]  C_msg<MSG-1 -> (C_msg'=C_msg+1);
 [messageC] C_msg<MSG-1 -> (C_msg'=C_msg+1) & (Comp'=true);
 [messageR] C_msg=MSG-1 -> (C_msg'=0) & (Comp'=false);

endmodule

rewards "Replacements"
  [messageR] true:  1;
endrewards
```

## F. PRISM Model for Hy/MB Key Update

```
ctmc
// time unit: 1 day

const int J; // threshold for number of joins
const int N=J; // threshold for number of leaves
const int M=J; // number of months between resets
const int k;
const double mean = 30*M;

const int    Max;       // Maximum number of devices
const double R_join;    // Rate of join per device
const double R_leave;   // Rate of leave per device
const double R_message; // Rate of message per device
const double P_comp;    // Risk of key leakage per device

module DEVICES
 Size: [0..Max] init Max;

 [join]     Size<Max  -> R_join*(Max-Size):     (Size'=Size+1);
 [joinR]    Size<Max  -> R_join*(Max-Size):     (Size'=Size+1);
 [leave]    Size>0    -> R_leave*(1-P_comp)*Size: (Size'=Size-1);
 [leaveC]   Size>0    -> R_leave*P_comp*Size:    (Size'=Size-1);
 [leaveR]   Size>0    -> R_leave*Size:          (Size'=Size-1);
 [message]  Size>0    -> R_message*(1-P_comp)*Size:          true;
 [messageC] Size>0    -> R_message*P_comp*Size:          true;
 [reset]    true      -> 1:                  true;
endmodule

module COORDINATOR
 Comp: bool init false;
 C_join: [0..J] init 0;
 C_leave: [0..N] init 0;
 i : [1..k+1];

 [join]   C_join<J-1 -> (C_join'=C_join+1);
 [joinR]  C_join=J-1-> (C_join'=0) & (C_leave'=0) & (i'=1) & (Comp'=false);
```

```
[leave]  C_leave<N-1 -> (C_leave'=C_leave+1);
[leaveC] C_leave<N-1 -> (C_leave'=C_leave+1) & (Comp'=true);
[leaveR] C_leave=N-1-> (C_join'=0) & (C_leave'=0) & (i'=1) & (Comp'=false);
[message] true  -> true;
[messageC] true -> (Comp'=true);
[]          i < k -> k/mean :      (i'=i+1);
[reset] i = k -> k/mean : (C_join'=0) & (C_leave'=0) & (i'=1) & (Comp'=false);
endmodule

rewards "Replacements"
  [joinR] true: 1;
  [leaveR] true: 1;
  [reset] true: 1;
endrewards
```

## G. CSL Formulae

```
const double T;

// Question 1:  Key compromise in the long run
S=? [ Comp ]

// Question 2: Number of key updates
R{"Replacements"}=? [ C<=30*T ]

// Question 3: Key compromise at monthly time instants
P=? [ F[30*T,30*T] Comp ]
```

## H. Input Parameters

We have presented all the input values for our analyses in Table 4. Further details on the parameters can be found in [11YNN+]. In addition, Gauss-Seidel method is used as the linear equation method, when necessary.

**Table 4: Input Values**

|  | LB | JB | JLB | TB | MB | Hy/MB |
|---|---|---|---|---|---|---|
| **MAX** | 50 | | | | | |
| **Rjoin** | 0.5 | | | | | |
| **Rleave** | 0.00274 | | | | | |
| **Rmessage** | 1 | | | | | |
| **Pcomp** | 0.0001 | | | | | |
| **k** | N/A | N/A | N/A | 100 | N/A | 100 |
| **Threshold Values** | 1, 2, 3, 4, 5 | 1, 2, 3, 4, 5 | 1, 2, 3, 4, 5 | 1, 2, 3, 4, 5 | 500, 1000, 1500, 2000, 2500 | 1, 2, 3, 4, 5 |
| **Threshold Unit** | Device | Device | Device | Month | Message | Device and Month |
| **Time Unit** | Day | | | | | |

# I. State Space

In this appendix, we present the state space information for the running example.

## I.1 State Space for LB

In Figure 22, we present the influence of the two main parameters: Network size (max), and Threshold on the state space. Further details on the exact numerical values for states as well as transitions can be found in Table 5 and Table 6.

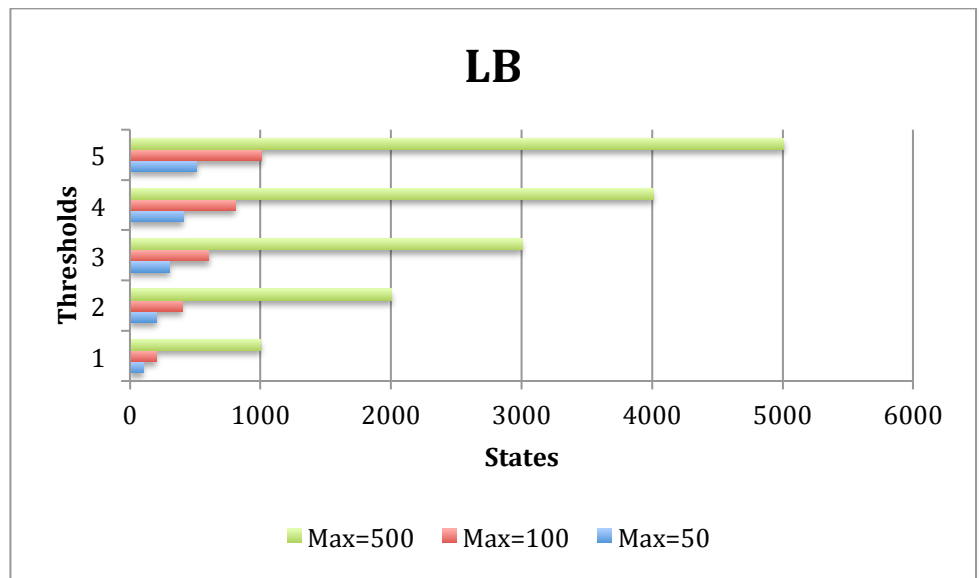

**Figure 22: Influence of the network size and the key update threshold on the state space, in LB.**

Table 5: Number of states in LB strategy.

|  |  | Max | | |
|---|---|---|---|---|
|  |  | 50 | 100 | 500 |
| Thresholds | 1 | 101 | 201 | 1001 |
|  | 2 | 203 | 403 | 2003 |
|  | 3 | 305 | 605 | 3005 |
|  | 4 | 407 | 807 | 4007 |
|  | 5 | 509 | 1009 | 5009 |

Table 6: Number of transitions in LB strategy.

|  |  | Max | | |
|---|---|---|---|---|
|  |  | 50 | 100 | 500 |
| Thresholds | 1 | 349 | 699 | 3499 |
|  | 2 | 749 | 1499 | 7499 |
|  | 3 | 1149 | 2299 | 11499 |
|  | 4 | 1549 | 3099 | 15499 |
|  | 5 | 1949 | 3899 | 19499 |

## I.2 State Space for JB

In Figure 23, we present the influence of the two main parameters: Network size (max), and Threshold on the state space. Further details on the exact numerical values for states as well as transitions can be found in Table 7 and Table 8.

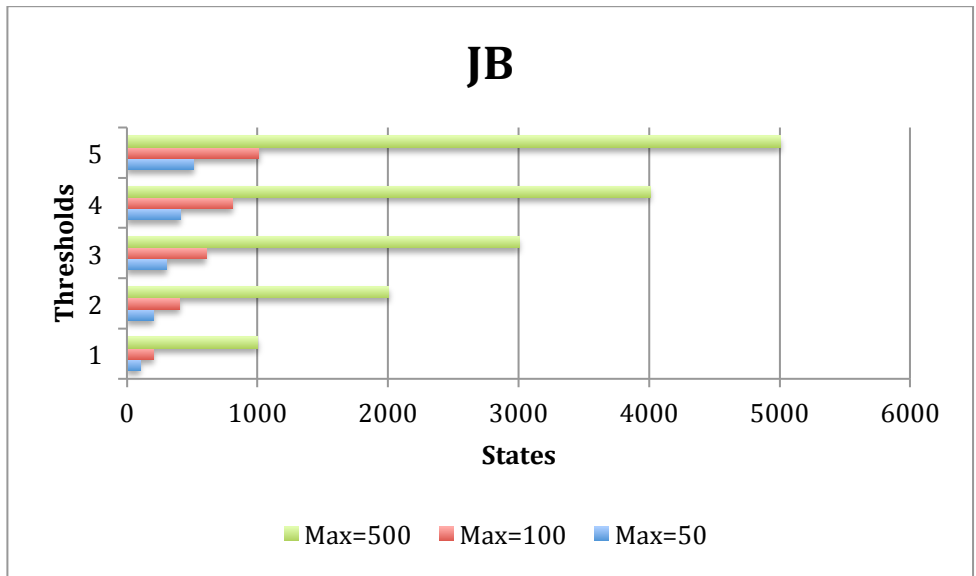

**Figure 23: Influence of the network size and the key update threshold on the state space, in JB.**

**Table 7: Number of states in JB strategy.**

|  |  | Max | | |
|---|---|---|---|---|
|  |  | 50 | 100 | 500 |
| Thresholds | 1 | 102 | 202 | 1002 |
|  | 2 | 204 | 404 | 2004 |
|  | 3 | 306 | 606 | 3006 |
|  | 4 | 408 | 808 | 4008 |
|  | 5 | 510 | 1010 | 5010 |

**Table 8: Number of transitions in JB strategy.**

|  |  | Max | | |
|---|---|---|---|---|
|  |  | 50 | 100 | 500 |
| Thresholds | 1 | 400 | 800 | 4000 |
|  | 2 | 800 | 1600 | 8000 |
|  | 3 | 1200 | 2400 | 12000 |
|  | 4 | 1600 | 3200 | 16000 |
|  | 5 | 2000 | 4000 | 20000 |

## I.3 State Space for JLB

In Figure 24, we present the influence of the two main parameters: Network size (max), and Threshold on the state space. Further details on the exact numerical values for states as well as transitions can be found in Table 9 and Table 10.

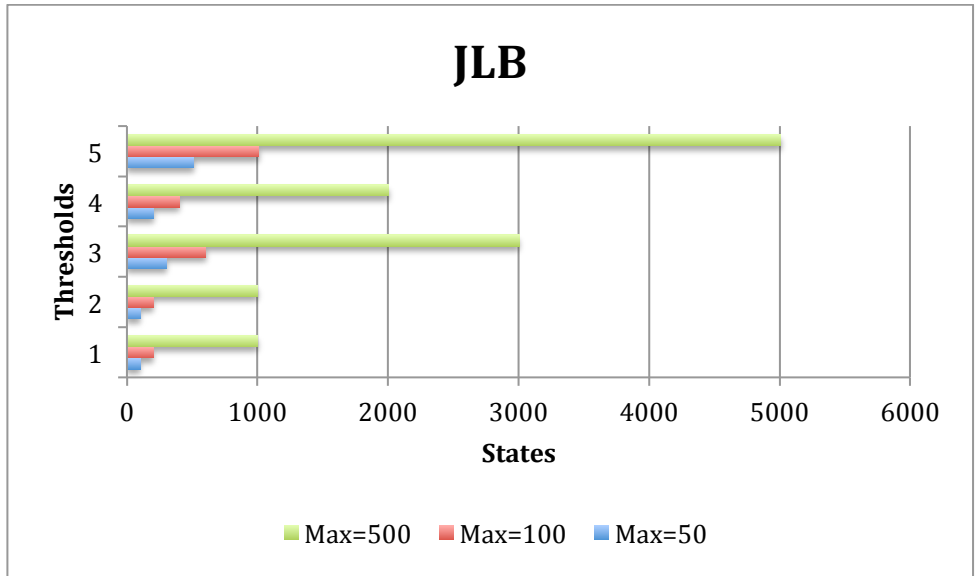

**Figure 24: Influence of the network size and the key update threshold on the state space, in JLB.**

Table 9: Number of states in JLB strategy.

|  |  | Max |  |  |
|---|---|---|---|---|
|  |  | 50 | 100 | 500 |
| Thresholds | 1 | 101 | 201 | 1001 |
|  | 2 | 101 | 201 | 1001 |
|  | 3 | 305 | 605 | 3005 |
|  | 4 | 203 | 403 | 2003 |
|  | 5 | 509 | 1009 | 5009 |

Table 10: Number of transitions in JLB strategy.

|  |  | Max |  |  |
|---|---|---|---|---|
|  |  | 50 | 100 | 500 |
| Thresholds | 1 | 349 | 699 | 3499 |
|  | 2 | 374 | 749 | 3749 |
|  | 3 | 1149 | 2299 | 11499 |
|  | 4 | 774 | 1549 | 7749 |
|  | 5 | 1949 | 3899 | 19499 |

## I.4 State Space for TB

In Figure 25, we present the influence of the two main parameters: Network size (max), and Threshold on the state space. Further details on the exact numerical values for states as well as transitions can be found in Table 11 and Table 12.

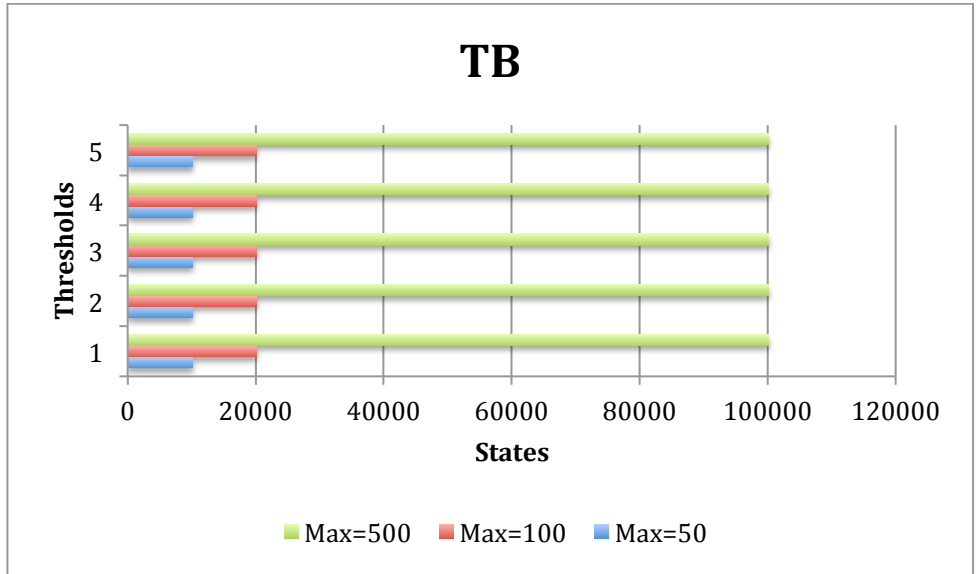

**Figure 25: Influence of the network size and the key update threshold on the state space, in TB.**

**Table 11: Number of states in TB strategy.**

|  |  | Max | | |
|---|---|---|---|---|
|  |  | 50 | 100 | 500 |
| Thresholds | 1 | 10200 | 20200 | 100200 |
|  | 2 | 10200 | 20200 | 100200 |
|  | 3 | 10200 | 20200 | 100200 |
|  | 4 | 10200 | 20200 | 100200 |
|  | 5 | 10200 | 20200 | 100200 |

**Table 12: Number of transitions in TB strategy.**

|  |  | Max | | |
|---|---|---|---|---|
|  |  | 50 | 100 | 500 |
| Thresholds | 1 | 50200 | 100200 | 500200 |
|  | 2 | 50200 | 100200 | 500200 |
|  | 3 | 50200 | 100200 | 500200 |
|  | 4 | 50200 | 100200 | 500200 |
|  | 5 | 50200 | 100200 | 500200 |

## I.5 State Space for MB

In Figure 26, we present the influence of the two main parameters: Network size (max), and Threshold on the state space. Further details on the exact numerical values for states as well as transitions can be found in Table 13 and Table 14.

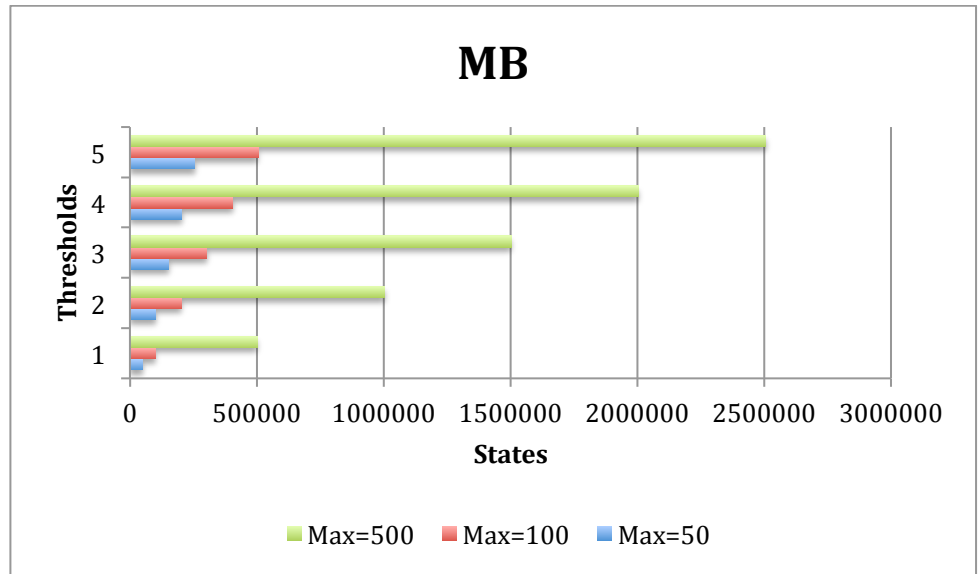

**Figure 26: Influence of the network size and the key update threshold on the state space, in MB.**

Table 13: Number of states in MB strategy.

|  |  | Max | | |
|---|---|---|---|---|
|  |  | 50 | 100 | 500 |
| Thresholds | 500 | 51000 | 101000 | 501000 |
|  | 1000 | 102000 | 202000 | 1002000 |
|  | 1500 | 153000 | 303000 | 1503000 |
|  | 2000 | 204000 | 404000 | 2004000 |
|  | 2500 | 255000 | 505000 | 2505000 |

Table 14: Number of transitions in MB strategy.

|  |  | Max | | |
|---|---|---|---|---|
|  |  | 50 | 100 | 500 |
| Thresholds | 500 | 199950 | 399900 | 1999500 |
|  | 1000 | 399950 | 799900 | 3999500 |
|  | 1500 | 599950 | 1199900 | 5999500 |
|  | 2000 | 799950 | 1599900 | 7999500 |
|  | 2500 | 999950 | 1999900 | 9999500 |

## I.6 State Space for Hy

In Figure 27, we present the influence of the two main parameters: Network size (max), and Threshold on the state space. Further details on the exact numerical values for states as well as transitions can be found in Table 15 and Table 16.

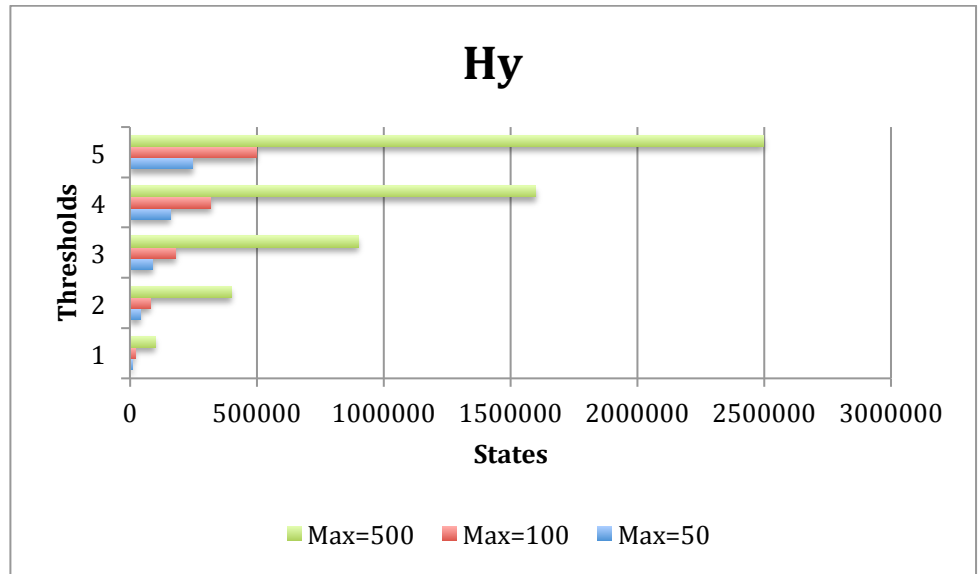

**Figure 27: Influence of the network size and the key update threshold on the state space, in Hy.**

**Table 15: Number of states in Hy strategy.**

|  |  | Max |  |  |
|---|---|---|---|---|
|  |  | 50 | 100 | 500 |
| Thresholds | 1 | 10100 | 20100 | 100100 |
|  | 2 | 40300 | 80300 | 400300 |
|  | 3 | 90100 | 180100 | 900100 |
|  | 4 | 159100 | 319100 | 1599100 |
|  | 5 | 246900 | 496900 | 2496900 |

In

**Table 16: Number of transitions in Hy strategy.**

|  |  | Max |  |  |
|---|---|---|---|---|
|  |  | 50 | 100 | 500 |
| Thresholds | 1 | 45000 | 90000 | 450000 |
|  | 2 | 189500 | 379500 | 1899500 |
|  | 3 | 431300 | 866300 | 4346300 |
|  | 4 | 768400 | 1548400 | 7788400 |
|  | 5 | 1198800 | 2423800 | 12223800 |